\documentclass[10pt,journal]{IEEEtran}
\usepackage{ifpdf}
\usepackage[hyphens]{url}
\usepackage{graphicx}
\usepackage{booktabs}
\usepackage{threeparttable}
\usepackage[cmex10]{amsmath}
\usepackage{mathtools} 
\usepackage{algorithmic}
\usepackage{array}
\usepackage{stfloats}
\usepackage[switch]{lineno}
\usepackage{color}
\usepackage{pifont}
\usepackage{hhline}
\usepackage[]{footmisc}
\usepackage{soul}
\usepackage{comment}
\usepackage[backend=bibtex,style=numeric,natbib=true,sorting=none,url=false,doi=false,isbn=false]{biblatex} 
\usepackage{lineno}
\newcommand{\bi}{\begin{itemize}}
\newcommand{\ei}{\end{itemize}}
\newcommand{\bd}{\begin{displaymath}}
\newcommand{\ed}{\end{displaymath}}
\newcommand{\be}{\begin{equation}}
\newcommand{\ee}{\end{equation}}
\newcommand{\bea}{\begin{eqnarray}}
\newcommand{\eea}{\end{eqnarray}}
\newcommand{\ba}{\begin{array}}
\newcommand{\ea}{\end{array}}
\newcommand{\bc}{\begin{center}}
\newcommand{\ec}{\end{center}}

\title{Physically interacting humans regulate muscle coactivation to improve visuo-haptic perception}

\author{Hendrik B\"orner$^{+,1}$, Gerolamo Carboni*$^{,+,2}$, Xiaoxiao Cheng$^{+,2}$,\\ Atsushi Takagi$^3$, Sandra Hirche$^{1}$, Satoshi Endo$^{1}$ and Etienne Burdet*$^{,2}$.

\thanks{+Equal contribution}
\thanks{*Corresponding authors:\{gerolamo.carboni16,e.burdet\}@imperial.ac.uk}
\thanks{This work was supported in part by the EC H2020 grants PH-CODING (FETOPEN 829186), CONBOTS (ICT 871803), REHYB (ICT 871767).}
\thanks{$^{1}$ Technical University of Munich, Electrical and Computer Engineering Department, München, 80333, Germany.
        }
\thanks{$^{2}$ Department of Bioengineering, Imperial College of Science, Technology and Medicine, London, SW7 2AZ, UK.
     }
\thanks{$^{3}$NTT Communication Science Laboratories, 3‐1 Morinosato Wakamiya, Atsugi, Kanagawa, 243-0198, Japan.
     }      
}

\begin{document}
\maketitle 

\begin{abstract}
When moving a piano or dancing tango with a partner, how should I control my arm muscles to best feel their movements and follow or guide them smoothly? Here we observe how physically connected pairs tracking a moving target with the arm modify muscle coactivation with their visual acuity and the partner's performance. They coactivate muscles to stiffen the arm when the partner's performance is worse, and relax with blurry visual feedback. Computational modelling show that this adaptive sensing property cannot be explained by movement error minimization proposed in earlier models. Instead, individuals skilfully control the arm’s stiffness to guide it on the planned motion while minimizing effort and extracting useful haptic information from the partner’s movement. The central nervous system regulates muscles' activation to guide motion with accurate task information from vision and haptics while minimizing the metabolic cost.
\end{abstract}

{\textit{New \& Noteworthy}:} Our results demonstrate (for the first time) that interacting humans inconspicuously modulate muscles' activation to extract accurate information about the common target while considering own and the partner's noise. A novel computational modelling was developed to decipher the underlying mechanism: muscle coactivation is adapted to combine haptic information from the interaction with the partner and own visual information in a stochastically optimal manner, thereby improving the prediction of the target position with minimal metabolic cost.

\section{Introduction}
\begin{figure}[tb]
\centering
\includegraphics[width=1\columnwidth]{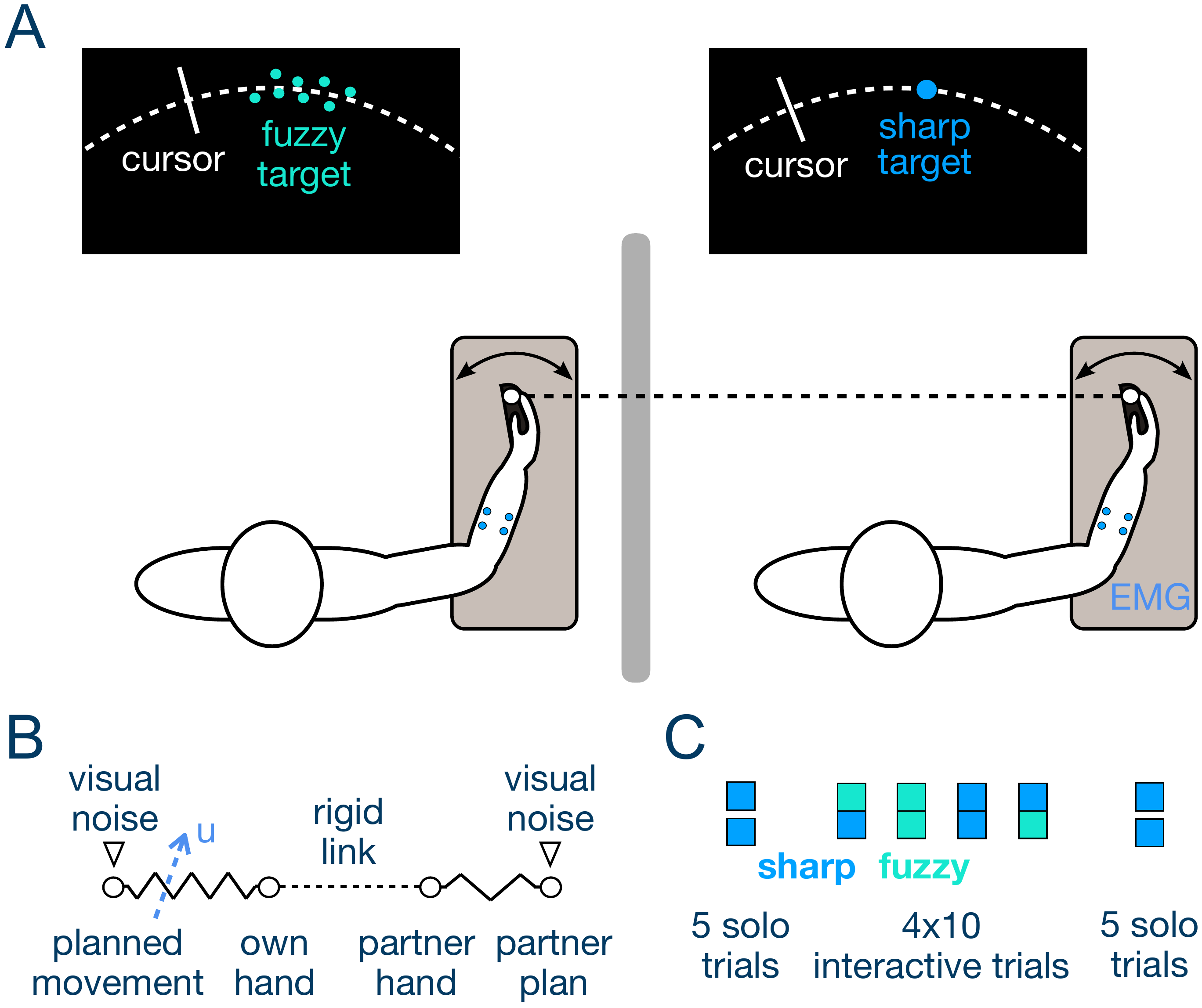}
\includegraphics[width=1\columnwidth]{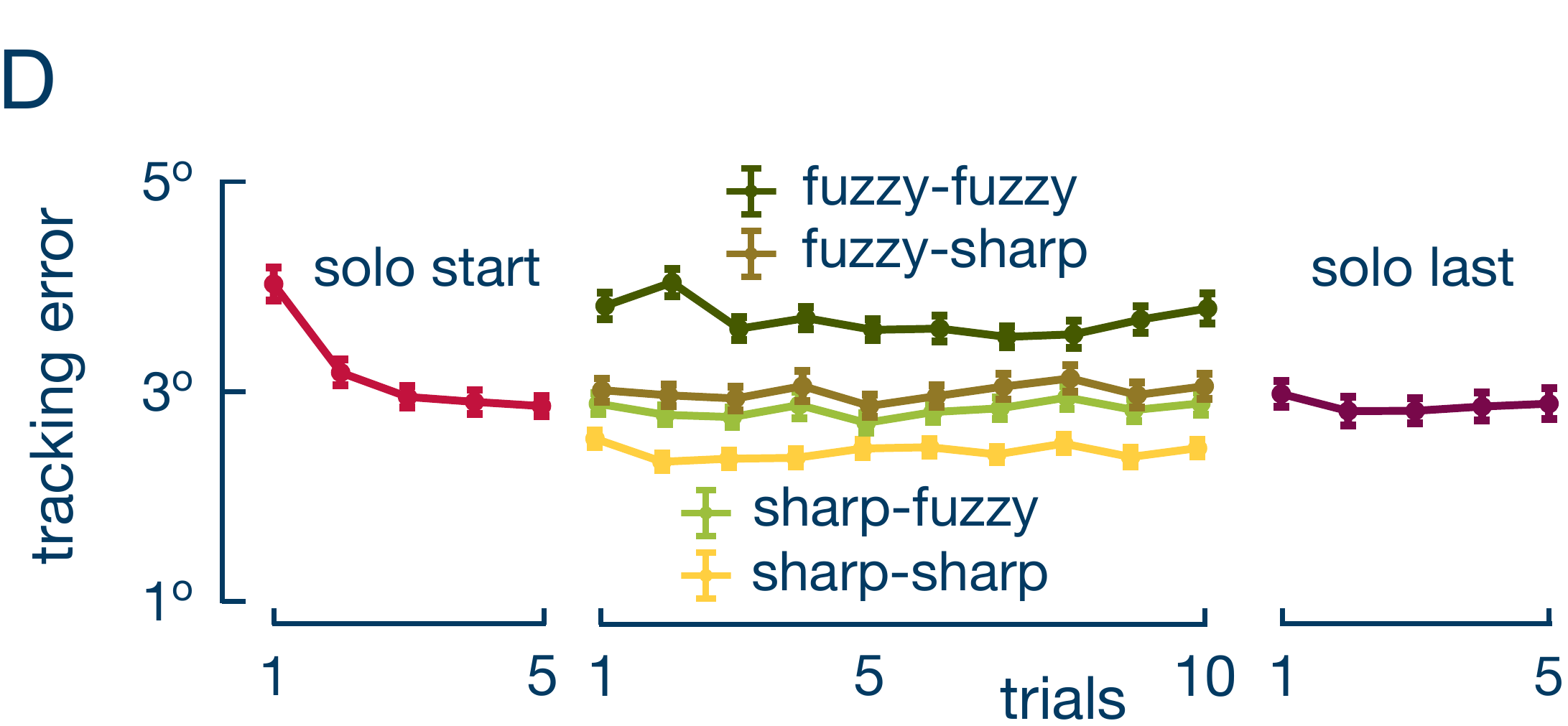}
\caption{Experiment to study how muscle coactivation varies during the joint tracking of two connected humans. (A) Two partners track the same randomly moving target with their wrist flexion-extension movement while being connected with a rigid virtual bar. Their wrist flexion/extension movement is recorded, as well as the myoelectrical activity of a flexor-extensor muscle pair. (B) Diagram of mechanical interaction with the partner and with own movement plan. The interaction with the partner's hand depends on the stiffness of the connection to their motion plan modulated by their coactivation $u$. Both own and the partner movement plans are affected by the respective visual noise. (C) Protocol of the experiment to study the effect of visual noise on either partner's performance and coactivation. (D) Evolution of group mean tracking error charted as a function of trials, where error bars represent one standard error. The error saturated in the initial solo trials, and increased with visual and haptic noise.}
\label{f:experiment}
\end{figure}

Human muscles are elastic elements that increase stiffness and shorten with activation \cite{Kirsch1994}. The central nervous system (CNS) regulates the limbs' stiffness by coordinating muscles' activation to shape the interaction with the environment \cite{Hogan1984a, Franklin2008}, but how this affects haptic sensing is not known. When two connected individuals carry out a task together (Fig.\,\ref{f:experiment}A), they exchange haptic information about their motion plan to combine with own visual information and improve their accuracy \cite{Takagi2017}. Critically, haptic information transferred by the mechanical connection is modulated by their muscle coactivation (Fig.\,\ref{f:experiment}B). Could individuals regulate their muscles' activation to adapt the limbs' stiffness and better sense the partner's movement? 

To understand how physically connected individuals control their arm coactivation, we observed 22 pairs of subjects or \textit{dyads} tracking a common randomly moving target using wrist flexion and extension (Fig.\,\ref{f:experiment}A). Studies on the adaptation to unpredictable force fields \cite{Takahashi2001, Burdet2001} suggest that muscle coactivation would increase with the magnitude of error to their motion plan \cite{Franklin2008} independent of its source. However, we hypothesized that interacting humans can adapt their muscle coactivation to their own sensorimotor noise and to haptic noise resulting from the interaction with the partner.

In our experiment, the visual feedback provided to the partners was manipulated to test this hypothesis. The target observed by each partner on their individual monitor was either \textit{sharp} (a 8\,mm large disk) or \textit{fuzzy} (a dynamic cloud of 8 normally distributed dots) as described in the Methods section. The tracking performance and wrist muscles coactivation of each partner was observed in four \textit{interaction conditions}: sharp (self) - fuzzy (partner) (SF), sharp - sharp (SS), fuzzy - sharp (FS) and fuzzy - fuzzy (FF).

The experiment was carried out as a within-subject design with the four interaction conditions randomly presented in a block of ten trials per condition (Fig.\,\ref{f:experiment}C). The forty interaction trials were preceded by five solo trials without interaction with the partner to learn the task, and followed with five solo trials. During the movement, the wrist flexion-extension angle as well as the muscle electromyographic activity (EMG) of the flexorcarpi radialis (FCR) and extensor carpi radialis longus (ECRL) were recorded from each participant. We analyzed the root mean square \textit{tracking error} to the target trajectory and the \textit{cocontraction} normalized to the torque as described in the Methods.

\section{Methods}
\label{sec:methods}
\subsection{Participants}
The experiment was approved by the Joint Research Compliance Office at Imperial College London. 44 participants without known sensorimotor impairments, aged 18\textendash37 years, including 16 females, were recruited. Each participant gave written informed consent prior to participation. 37 participants were right-handed and 5 left-handed, as was assessed using the Edinburgh Handedness Inventory \cite{Oldfield1971}. The participants carried out the experiment in pairs or \textit{dyads}, with 14 male-male dyads and 8 female-female dyads. 

\subsection{Experimental setup}
\label{sec:setup}
The two partners of each dyad were seated comfortably on height-adjustable chairs, next to the Hi5 dual robotic interface \cite{Melendez-Calderon2011}. They held their respective handle with the right wrist and received visual feedback of the flexion/extension movement on a personal monitor (Fig.\ref{f:experiment}A). No visual feedback of the partner's position was available as the two participants were separated by a curtain, and they were instructed not to speak to each other during the experiment.

Each Hi5 handle is connected to a current-controlled DC motor (MSS8, Mavilor) that can exert torques of up to 15\,Nm, and is equipped with a differential encoder (RI 58-O, Hengstler) to measure the wrist angle and a (TRT-100, Transducer Technologies) sensor to measure the exerted torque in the range [0,11.29]\,Nm. The two handles are controlled at 1\,kHz using Labview Real-Time v14.0 (National Instruments) and a data acquisition board (DAQ-PCI-6221, National Instruments), while the data was recorded at 100\,Hz. 

The activation of two antagonist wrist muscles, the flexor carpi radialis (FCR) and extensor carpi radialis longus (ECRL) were recorded during the movement from each participant. Electromyographic (EMG) signals were measured with surface electrodes using the medically certified g.Tec's g.LADYBird\&g.GAMMABox\&g.BSamp system. The EMG data was recorded at 100\,Hz.

\subsection{Tracking task}
The two partners were required to track the same \textit{visual target} (in degrees) moving on their respective monitor with
\begin{eqnarray}
q^*(t) \!\!\!&\equiv&\!\!\!  \, 18.5 \, \sin \! \left(2.031\,t^* \!\right)  \, \sin\!\left(1.093\,t^*\!\right) \nonumber \\
t^* \!\!\!&\equiv&\!\!\! t + t_0 \,\,, \quad 0\leq t\leq 20\,s
\label{eq:b_traj}
\end{eqnarray}
as accurately as possible using flexion-extension movements (Fig.\ref{f:experiment}A). To prevent the participants from memorizing the target's motion, $t^*$ started in each trial from a randomly selected zero $\{t_{0} \in [0, 20]\,\mathrm{s}\ |\ q^*(t_{0}) \equiv 0\}$ of the multi-sine function. The respective \textit{tracking error}
\be
e \equiv \, \left(\!\frac{1}{T} \! \int_0^T \!\!\! \left[q^*\!(t) - q(t) \right]^2 dt \!\right)^\frac{1}{2} \!\!,\, \,\,\, T \equiv 20\,s
\label{e:error}
\ee
was displayed at the end of each 20\,s long trial.

After each trial, the target disappeared and the participants had to place their respective cursor on the starting position at the center of the screen. The next trial then started after a 5\,s rest period and a 3\,s countdown. The initialization of next trial started when both participants placed their wrist on the starting position, so that each participant could take a break at will in between trials, by keeping the cursor away from the center of the screen.

\subsection{Experimental conditions and protocol}
In \textit{solo trials}, the two partners moved the wrist independently to each other. In \textit{interactive trials}, the partners' wrists were connected by a stiff virtual spring with torque (in Nm)
\be
\tau(t) = \,17.2 \,[q_p(t)-q_o(t)] \, ,
\ee
where $q_o$ and $q_p$ (in radian) denote own and the partner's wrist angles, respectively. As the tracking errors of the two partners of a dyad were strongly correlated (r(20)=0.95, p$<$0.0005), the average tracking error between them was used in the data analysis.

The interaction trials were carried out under two different visual feedback conditions. In the \textit{sharp condition} the target was displayed as a 8\,mm diameter disk. In the \textit{fuzzy condition} the target trajectory was displayed as a ``cloud'' of eight normally distributed dots around the target. The cloud of dots were defined by three parameters, randomly picked from independent Gaussian distributions: the vertical distance to the target position $\eta \in$ N(0,\,15\,mm), the angular distance to the target position $\eta_q \in$ N(0,\,4.58$^\circ$), and the angular velocity $\eta_{\dot{q}} \in$ N(0,\,4.01$^\circ$/s). Each of the eight dots was sequentially replaced every 100\,ms.

A calibration of the measured EMG (described in Section \ref{sec:calibration}) was first carried out to map the raw EMG signal (in mV) to a corresponding torque value (in Nm), so that the activity of each participant's flexor and extensor's can be compared and combined in the data analysis. After this calibration, the participants carried out 5 initial solo trials to learn the tracking task and the dynamics of the wrist interface. This was followed by 4 blocks of 10 interaction trials, each with one of the four different noise conditions \{fuzzy(self)-sharp(partner): FS, SF, SS, FF\} presented in a random order, followed by 5 control solo trials (Fig.\ref{f:experiment}B). The participants were informed when an experimental condition would be changed but not which condition would be encountered in the next trials.

\subsection{Muscle activation calibration and cocontraction calculation}
\label{sec:calibration}
The participants placed their wrist in the most comfortable middle posture, set to 0$^\circ$. Constrained at that posture, they were then instructed to sequentially {(\it i}) flex or extend the wrist to exert a torque, or {(\it ii}) maximally co-contract in order to keep the wrist position stable during a 3\,Hz sinusoidal positional disturbance of 10$^\circ$ amplitude. Each phase was 4\,s long and was followed by a 5\,s rest period to avoid fatigue. The latter period was used as a reference activity in the relaxed condition. This procedure was repeated four times at flexion/extension torque levels of \{1,2,3,4\}\,Nm and \{-1,-2,-3,-4\}\,Nm, respectively. For each participant, the recorded muscle activity was then linearly regressed against the torque values to estimate the relationship between them. The raw EMG signal was first high-pass filtered at 20\,Hz using a second-order Butterworth filter to remove drifts in the EMG signal. This was then rectified and passed through a low-pass second-order Butterworth filter with a 5\,Hz cut-off frequency to obtain the envelope of the EMG activity.

The torque of the flexor muscle could then be modelled from the envelope of the EMG activity $u_f$ as
\begin{equation}
	\begin{aligned}
		\tau_f(t) = \, \alpha_0 \, u_f(t) \, + \, \alpha_1 \,, \quad \alpha_0, \alpha_1 > 0 \,,
	\end{aligned}
	\label{eq:regression}
\end{equation}
and similarly for the torque of the extensor muscle $\tau_e$. \textit{Muscle coactivation} was then computed as
\begin{equation}
	\begin{aligned}
		u(t) \equiv \, \textup{min}\{\tau_f(t),\tau_e(t)\}\,.
	\end{aligned}
	\label{eq:cocon}
\end{equation}
The average cocontraction over all participants (as shown in Fig.\ref{f:cocontraction}) was computed from each participant's normalised cocontraction, calculated as  
\be
u_n \equiv \frac{\overline{u} - \overline{u}_{min}}{\overline{u}_{max}-\overline{u}_{min}}\,, \quad 
\overline{u} \equiv \, \frac{1}{T} \!\int_{0}^{T} \!\!\!\! u(t) \, dt \,, \quad T \equiv 20\,s
\ee
with $\overline{u}_{min}$ and $\overline{u}_{max}$ the minimum and maximum of the means of all trials of the specific participant.

\subsection{Simulation of tracking error minimization (TEM)}
For each of the 4 noise conditions $ij$\,$\in$\,\{SS,\,SF,\,FS,\,FF\}, the initial cocontraction level $\{\hat{u}_{ij}(1)\}$ was set as the initial experimental value $\{u_{ij}(1)\}$. Then, by using the respective trial-by-trial tracking error $\{e_{ij}(k)\}, k=1,...,10$ from the experiment, the adaptation parameters $\alpha,\gamma$ in the computational model of eq.(\ref{eq_Vshape}) were computed to minimize the error between the learned cocontraction values after 9 iterations $\{\hat{u}_{ij}(10)\}$ and the corresponding data $\{u_{ij}(10)\}$ in last experiment's trial:
\be
(\alpha^*\!, \gamma^*) \equiv \, \underset{\alpha, \gamma}{\arg\min}
\bigg\{ \sum_{i,j}\left[\hat{u}_{ij}(10)-u_{ij}(10) \right]^2 \bigg\} \, .
\ee
The parameters $\alpha^* \equiv 0.5, \gamma^* \equiv 0.06$ were found by using a grid search with a step $0.01$ in the range $[0,2]\times[0,1.5]$.

\subsection{Simulation of optimal information and effort (OIE)}
A gradient descent optimisation was used to minimize the prediction error and effort
\be
V(u) = E(u) + \frac{1}{2}\gamma u^2\, , \quad E(u) \equiv \, \frac{\sigma_o^2(u) \, \sigma_p^2}{\sigma_o^2(u) + \sigma_p^2}\,.
\ee
Muscle cocontraction was updated trial after trial using:
\bea
u_{new} \!\!\!&=&\!\!\! u - \frac{dV(u)}{du} \,= \,- \frac{dE(u)}{du} \,+\, (1-\gamma)\,u \,, \nonumber \\
-\frac{dE(u)}{du} \!\!\!& = &\!\!\! \left[ \frac{\sigma_{p}^2}{\sigma_{o}^2 (u) +\sigma_{p}^2}\right]^2 \left[-\frac{d \sigma_{o}^2(u)}{du} \right] >0 \,.
\eea
The target tracking arises from the internal guidance to the planned motion and the mechanical connection with the partner, both being subjected to the noise in the respective visual feedback (Fig.\,\ref{f:experiment}C). How should the deviation $\sigma_o$ be modelled? Firstly, let $\sigma_{vo}$ describe the tracking deviation of own wrist movement due to the target cloud. And also, the wrist's compliance influences the tracking performance and brings in noise in the planned movement \cite{Takagi2018}. Assuming that these two effects are independent and that the wrist’s viscoelasticity results in zero mean noise with deviation $\sigma_{\kappa o}(u)$, the deviation in the wrist can be calculated as
\be \label{eq_HNModel}
\sigma^2_o(u) = \, \sigma^2_{vo} \, + \, \sigma^2_{\kappa o}(u) \,.
\ee

\begin{figure}[tb]
\begin{centering}
\includegraphics[width=1\columnwidth]{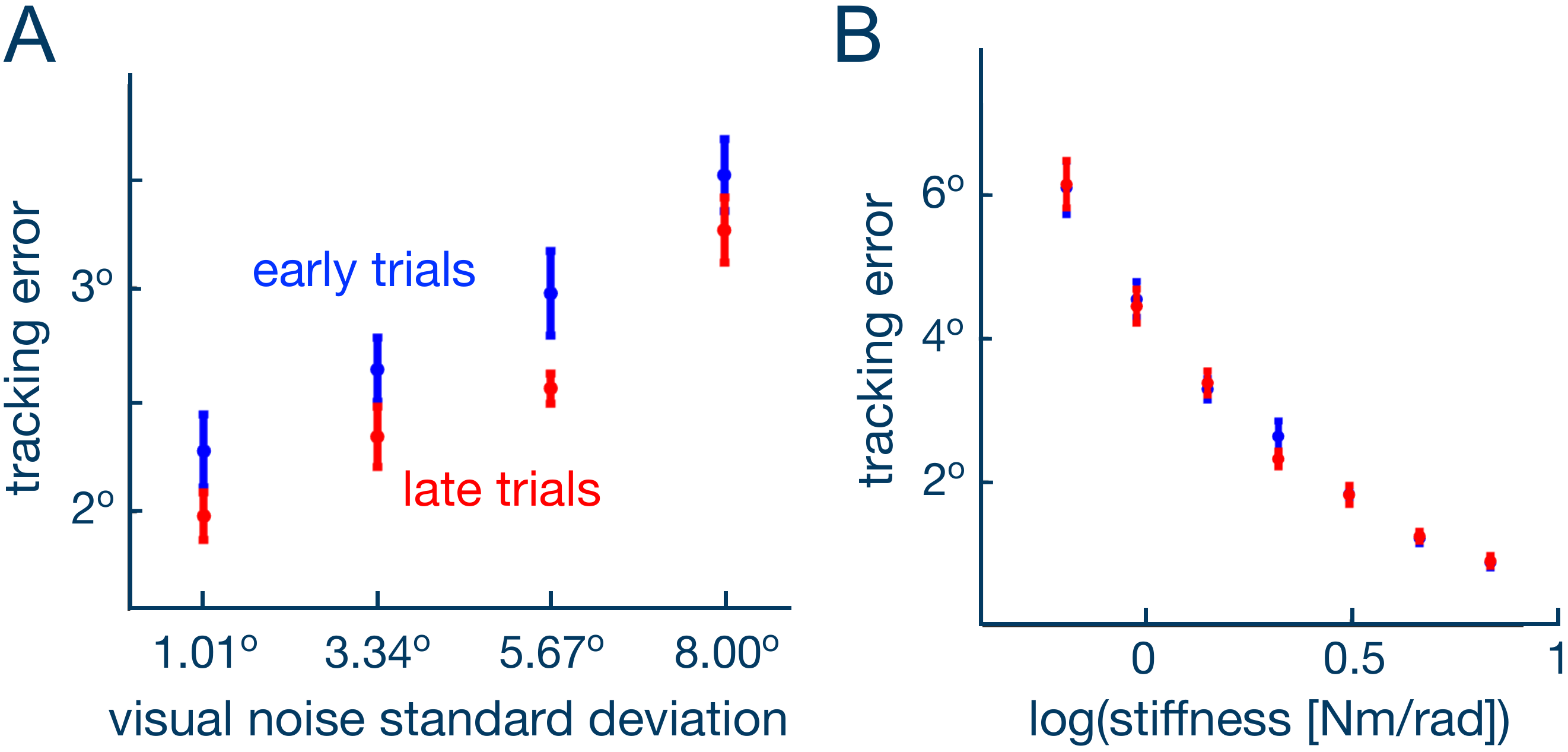}
\par\end{centering}
\caption{\label{f:control} 
Results of experiments to isolate the effect of visual feedback and haptic guidance on the tracking performance. (A) In the visual noise experiment, the tracking error linearly grows with visual noise, and decreased across all noise levels after learning. (B) In the experiment with haptic feedback only, the tracking error decreases with the log of the elasticity. This relationship did not change with practice.}
\end{figure}

To find the relation between the wrist's and the deviation $\sigma^2_{\kappa o}$, a \textit{haptic experiment} was first carried out to measure how the wrist compliance affects the tracking performance. In this control experiment, no visual feedback was provided to the participants while the wrist was connected to the target reference trajectory with 7 selected levels of elasticity. Eight participants not involved in the main experiments (7 right-handed, 26.75$\pm$1.28 years old, including 4 females) were recruited. The experiment was structured in 5 blocks of 7 elasticity $\kappa \in \{0.011, 0.016, 0.025, 0.037, 0.055, 0.081, 0.120\}$\,Nm/$^\circ$ presented in random order. In the interaction phase totalling 35 trials, the subjects experienced an elastic force to the target. As expected, the tracking error decreased with increasing elasticity (Fig.\ref{f:control}B), which was modelled as an exponential function
\be \label{eq_EToCC}
e(\kappa) = \,\, \alpha_h \, e^{-\beta_h \kappa} + \, \gamma_h\,,
\ee
where $\alpha_h$, $\beta_h$, $\gamma_h$ were identified from a least-square fit.

A \textit{visual experiment} was then carried out to relate the effect of the wrist elasticity to visual noise i.e. the deviation of the target cloud, by evaluating the influence of the visual noise on the tracking error. No elastic force was present in this control experiment. Another eight right-handed participants (25.01$\pm$0.53 years old, including 2 females) participated in this study. The results show that the tracking error is linearly correlated with the visual noise imposed on the target (Fig.\ref{f:control}A, \cite{Takagi2019}). This was modelled as
\be \label{eq_ETosigmav}
e(\sigma_v) = \alpha_{v} + \beta_{v} \, \sigma_v\,,
\ee 
where $\alpha_{v}, \beta_{v}$ were identified from a least-squares linear regression with data from late trials.

The deviation of the connection compliance could then be transformed to its equivalent value of visual noise deviation by setting $e(\kappa)\equiv e(\sigma_v)$, yielding
\be \label{eq_sigmac_exp}
\sigma_{\kappa o}(u) = \xi_0 \,+\, \xi_1 \,e^{-\beta_\kappa u} \quad \xi_0, \,\xi_1, \,\beta_\kappa >0
\ee
with $\xi_0$\,=\,5.18, $\xi_1$\,=\,49.65, $\beta_{\kappa}$\,=\,6.11. 

In the main experiment, considering the relationship between the deviation $\sigma_{ko}$ and the wrist's viscoelasticity, the visual noise deviation and the partner's noise deviation each has two values, resulting in four parameters to identify:
$\{\sigma_{vo}^{(0)}, \sigma_{vo}^{(n)}, \sigma_p^{(0)}, \sigma_p^{(n)} \}$, where $(0)$ represents the clean target while $(n)$ corresponds to the cloudy target. These parameters, used in the noise models of eqs.(\ref{eq_HNModel}, \ref{eq_sigmac_exp}), were computed by minimizing the variation of the cost derivative:
\bea \label{eq_DiffCost}
&\left(\sigma_{vo}^{(0)*}, \sigma_{vo}^{(n)*}\!,\sigma_{p}^{(0)*}\!, \sigma_{p}^{(n)*}\right) \equiv \\ & \underset{\sigma_{vo}^{(0)}, \sigma_{vo}^{(n)}, \sigma_{p}^{(0)}, \sigma_{p}^{(n)}}{\arg\min} \bigg\{ \displaystyle \sum_{i,j} \!\left[ \frac{\partial V}{\partial u} \left(u_{ij}(10), \, \sigma_{vo}^{(i)},\, \sigma_{p}^{(j)}\right) \right]^2 \!\bigg\}
\nonumber
\eea
Using the collected cocontraction data $\{u_{ij}(10)\}$, a grid search for $(\sigma_{vo}^{(0)}, \sigma_{vo}^{(n)}, \sigma_{p}^{(0)}, \sigma_{p}^{(n)})$ in $[0,10]\times[0,20]\times[0,10]\times[0,20]$ with step $0.2$ yields $\sigma_{vo}^{(0)*} = 10$, $\sigma_{vo}^{(n)*} = 18.8$, $\sigma_{p}^{(0)*}=5.2$, $\sigma_{p}^{(n)*}=6$, where for each gridpoint $\gamma^*=0.65$ was the solution of
\be
0 \equiv \, \frac{d}{d\gamma} \!\left( \displaystyle \sum_{i,j} \!\left[ \frac{\partial V}{\partial u} (u_{ij}(10), \sigma_{vo}^{(i)}, \sigma_{p}^{(j)})\right]^2 \right).
\ee

\section{Experimental results}
To evaluate the short-term adaptation within each condition, statistics were carried out on the average measurements from the first and second half of trials or epochs. As the two partners of a dyad are rigidly connected, the tracking error by trial was analysed per dyad using a two-way repeated-measures ANOVA with noise condition \{SS, FF, SF $\equiv$ FS\} and epoch as the factors. Statistical significance was detected at 5\% with Bonferroni correction for all post-hoc comparisons. As muscle cocontraction was modulated by each partner of a dyad, it was analysed individually wherein the partner's visual noise was perceived as haptic noise. Thus, a three-way repeated-measures ANOVA with visual noise, haptic noise and epoch as the factors was used in the analysis of the cocontraction level.

We see in Fig.\,\ref{f:experiment}D that the error had decreased by the last of the initial solo trials to the same degree as the average of the last solo trials (paired-sample t-test, t(21)=0.354, p=0.73). This indicates stable performance to analyse the different interaction conditions. The ANOVA of error in these conditions indicated that the magnitude of the tracking error depended on the noise level (F(1,21) = 91.95, p$<$0.001, $\eta^2_p$ = 0.81). Post-hoc comparisons showed that the tracking error in the mixed noise condition \{SF\,$\equiv$\,FS\} was greater than in the SS condition (p$<$0.001) and smaller than in the FF condition (p$<$0.001). The tracking error remained at a similar level between the first and the second epochs~(p=0.64), and there was no interaction effect between the noise level and epoch~(p=0.17).

\begin{figure}[!tb]
\centering
\includegraphics[width=0.84
\columnwidth]{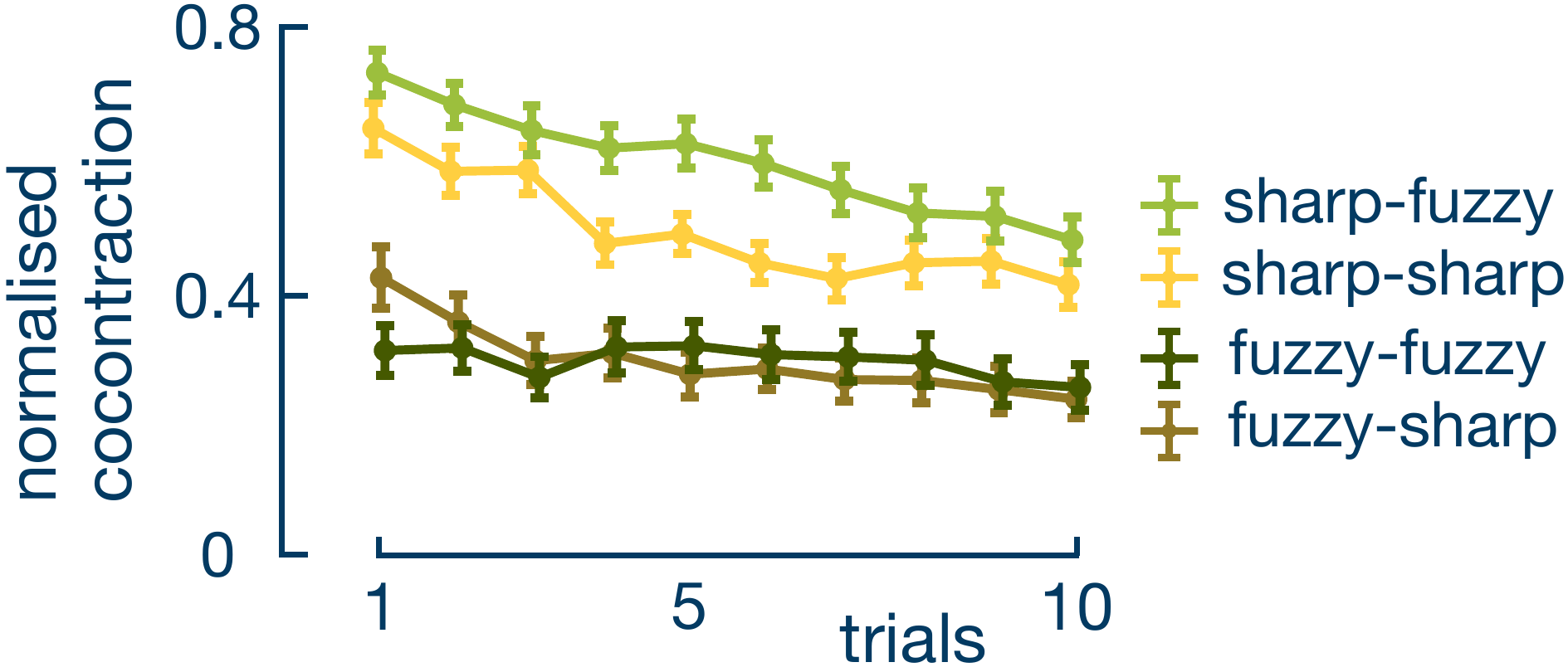}
\caption{Evolution of normalized cocontraction as a function of trials, where error bars represent one standard error. While visual noise increases the cocontraction by a larger margin than the haptic noise.}
\label{f:cocontraction} 
\end{figure}
The cocontraction level decreased with the epoch~(F(1,43)=53.58, p$<$0.0005, $\eta^2_p$ = 0.56), similar to what was observed during the learning of force fields \cite{Thoroughman1999, Franklin2008} (Fig.\ref{f:cocontraction}). 
Importantly, the cocontraction decreased with a larger level of own visual noise (F(1,43)=85.91, p$<$0.0005, $\eta^2_p$ = 0.67) while it increased with haptic noise from the interaction with the partner (F(1,43) = 5.53, p$<$0.03, $\eta^2_p$ = 0.11). Post-hoc comparisons confirmed that all differences between the combinations of the visual and haptic noises were significant with the exception of FS vs. FF (p = 0.99). 
These results demonstrate that \textit{during interaction the CNS spontaneously regulated muscle cocontraction considering the level of the visual noise on one's own and the partner's targets}, in agreement with our hypothesis.

\section{Computational modeling}
\begin{figure}[tb]
\centering
\includegraphics[width=\columnwidth]{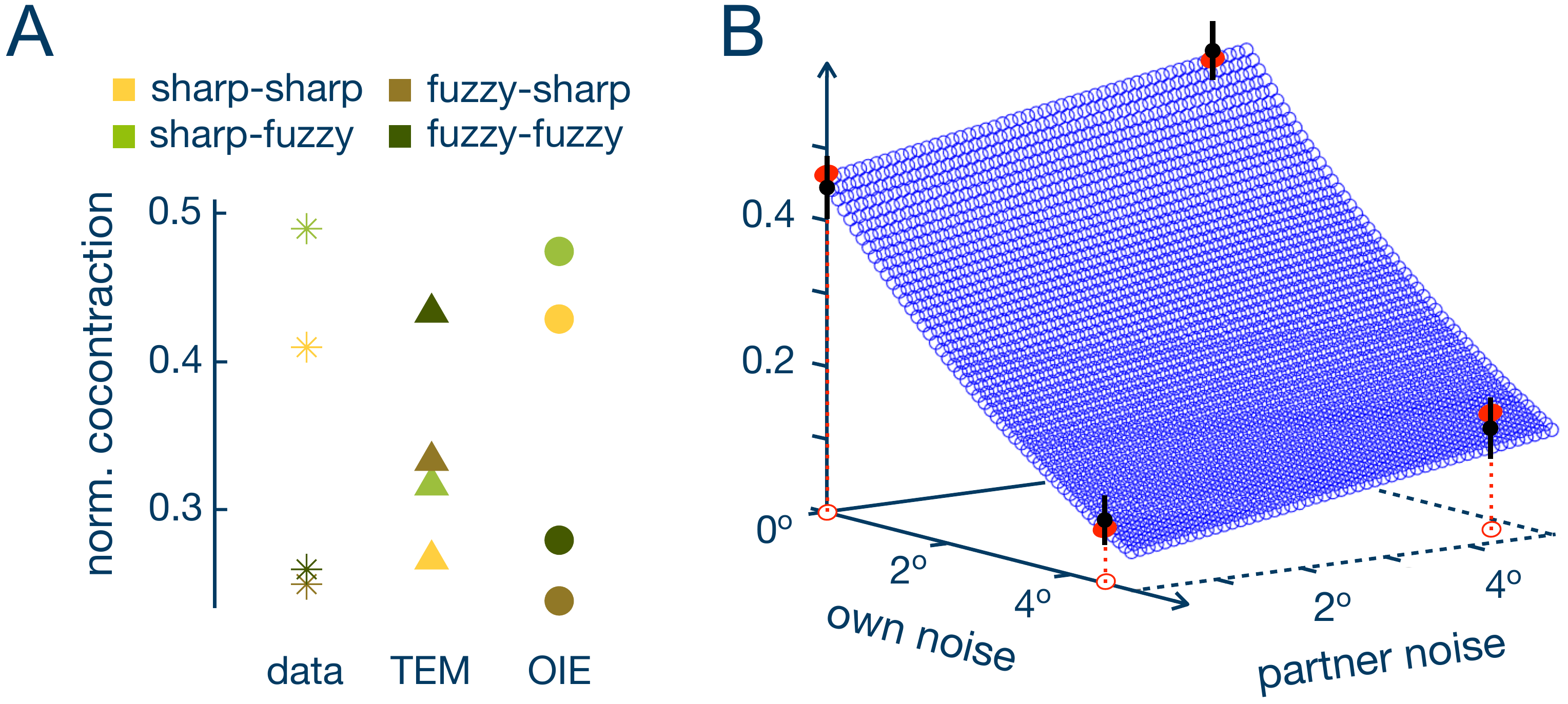}
\caption[]
{Results of computational modeling the coactivation adaptation to own and partner noise. (A) Comparison of coactivation observed during the experiment and predicted by the three models described in the text. The TEM model predicts a different modulation of coactivation with varying noise conditions as in the data, the PNM model predicts only the modulation with own visual noise. (B) The OIE model prediction (red ovals) exhibits a similar decrease of coactivation with own noise and increase with partner noise as in the data (black disks with standard error bars).}
\label{f:model}
\end{figure}

To clarify the coactivation adaptation mechanism during interaction, we first tested the computational model of \citep{Franklin2008} that explains the motor learning in novel force fields. In this model, the coactivation $u$ increases with each new trial to minimize tracking error $e$, and decreases to minimize effort, according to
\be \label{eq_Vshape}
u_{new} \equiv \, \alpha \, e + (1-\gamma)\,u \,, \quad \alpha,\gamma > 0 \, .
\ee
Simulation of the learning during the ten trials of each condition with this \textit{tracking error minimization} (TEM) model predicted cocontraction at a level increasing with the corresponding tracking error (Fig.\,\ref{f:model}A). This prediction is qualitatively different from the data, such as larger coactivation in the fuzzy relative to the sharp visual feedback condition (e.g. compare the FF and SS conditions in Fig.\,\ref{f:model}A). Therefore, the TEM model cannot explain the adaptation in the coactivation during interaction.

The hand's movement depends on the guidance towards the planned movement and on the connection to the partner (Fig.\,\ref{f:experiment}B). As the stiffness of the guidance increases with own coactivation \cite{Hogan1984a}, it is in principle possible to weight these two influences. Coactivation should decrease to lower the guidance to the planned movement when it is affected by visual noise. Conversely, the guidance to the planned motion should increase to counteract the effect of haptic noise when the partner receives noisy visual feedback. Therefore, the coactivation may depend both on the statistical information determining the quality of the planned motion, which relies primarily on vision, and on the partner's accuracy in tracking the common target. 

We thus propose that coactivation is modulated to maximise information from visual information and haptic information from the interaction with the partner. We introduce the \textit{optimal information and effort} (OIE) model that addresses the tradeoff between motion guidance and interaction noise attenuation, by selecting coactivation $u$ to minimize the \textit{prediction error}
\be
E(u) \equiv \, \frac{\sigma_o^2(u) \, \sigma_p^2}{\sigma_o^2(u) + \sigma_p^2}
\label{e:predictionError}
\ee
and \textit{metabolic cost} $u^2$, where $\sigma_o(u)$ results from the effect of own visual noise on the arm movement and $\sigma_p$ from the interaction with the partner. This optimization can be carried out through gradient descent minimization:
\be
u_{new} \equiv \,- \frac{dE(u)}{du} \,+\, (1-\gamma)\,u\,, \quad \gamma>0\,.
\label{e:OIE}
\ee
As can be seen in Fig.\,\ref{f:model}A, the OIE model predicts the modulation of coactivation with both own visual noise and haptic noise from the partner as observed in the data, in contrast to the TEM model.

\section{Discussion}
These experimental and computational results demonstrate that interacting humans inconspicuously modulate their muscles' activation to extract accurate information about the common task target, considering own and the partner's noise. While it has been known that muscles' activation is adapted to shape the mechanical interaction with the environment \cite{Hogan1984a, Burdet2001}, our results reveal that the CNS can regulate the limbs' viscoelasticity to extract optimal sensory information from the environment. Not only do individuals share haptic information to extract each other's motion plan \cite{Takagi2017}, but they further learn muscles' activation to improve this estimation.

These results could not be explained by a previous model of coactivation adaptation which considers only the error in the task performance \cite{Franklin2008, Theodorou2010, Li2018a}. However, the observed coactivation changes with both own and the partner's noise were well predicted by the OIE model introduced in this paper. The OIE adapts coactivation to maximize information from vision and haptics arising from the interaction with the partner, and minimize energy by reducing superfluous coactivation. This simple principle skilfully regulates coactivation to extract maximum sensory information while exploiting the guidance potential from the partner: Coactivation decreases to rely more on the partner guidance when vision is fuzzy, and increases when the interaction with the partner is noisy.

The OIE model, specifying how the CNS adapts coactivation to minimize prediction error and energy, extends previous work on motor learning and adaptation. While the models in  \cite{Herzfeld2014, Takiyama2015} could determine the motion kinematics in the next trial from the movement error in previous trials, this new model also considers the limbs' neuromechanics and can thus predict the interaction force and the subsequent muscle activity during motion. The OIE also extends optimal and nonlinear adaptive control models \cite{Franklin2008, Nelson1983, Harris1998, Todorov2002, Li2018a} by considering the consequence of action on the acquired sensory information from the environment, closing the loop between the sensory and motor actions. This adaptive sensing mechanism may give rise to interactive robots that can modify their rigidity to optimally sense their user and best assist them.

\printbibliography

\section*{Contribution Statement}
The contributions are as follows:
\begin{itemize}  
\item HB, GC, SH, EB conceptualised the experiment. 
\item HB, GC, carried out the experiment. 
\item HB, GC, AT, SE, EB performed data and statistical analysis. 
\item HB, GC, XC, EB developed the computational modeling.
\item All authors have edited the text and agree with its content.
\end{itemize}

\textbf{Competing interests:} The authors declare no competing interests.


\end{document}